\documentclass[prl,aps,amssymb,amsmath]{article}
\usepackage{amssymb,amsmath,tikz}
\usepackage{amsthm}
\usepackage{mathrsfs}
\usepackage{pifont}

\usepackage{amsfonts,amsmath,amssymb}
\usepackage[none]{hyphenat}
\usepackage{fancyhdr}
\usepackage{graphicx}
\usepackage{float}
\usepackage[nottoc,notlot,notlof]{tocbibind}
\usepackage{amssymb,amsmath,tikz}
\usepackage{pgfplots} 
\usepackage{amsthm}
\usepackage{mathrsfs}
\usepackage{pifont}
\usepackage{slashed}
\usepackage{mathtools}
\usepackage{cancel}

\showoutput
\DeclareMathAlphabet{\mathpzc}{OT1}{pzc}{m}{it}

\DeclareMathOperator*{\supp}{supp}

\DeclareMathOperator*{\p}{p}
\DeclareMathOperator*{\x}{x}

\newcommand*{\ud}{\mathrm{\,d}}

\theoremstyle{plain}

\newtheorem*{twr*}{Theorem}
\newtheorem*{lem*}{Lemma}

\newtheorem*{defin*}{Definition}

\newtheorem*{rem*}{Remark}

\newtheorem{cor*}{Corollary}

\newtheorem*{notn*}{Notation}
\newtheorem*{wiener-ito*}{Wiener-It\^o-Segal Decomposition}
\newtheorem*{prop*}{Proposition}

\DeclareMathAlphabet{\mathpzc}{OT1}{pzc}{m}{it}

\showoutput
\theoremstyle{plain}

\begin{document}
\title{ {\bf Positivity of the Invariant Kernel Underlying Quantum Theory of the Coulomb Field}}
\author{Jaros{\l}aw Wawrzycki
\\
Institute of Nuclear Physics of PAS, ul. Radzikowskiego 152, 
\\31-342 Krak\'ow, Poland}
\maketitle

\vspace{1cm}

\begin{abstract}
We present a proof of positivity of an invariant kernel, 
which is of basic importance for the Staruszkiewicz theory of the quantum Coulomb field. 
Presented proof of positivity is independent of the Staruszkiewicz theory 
and is based on the classical Schoenberg's theorem for conditionally
negative definite functions, as well as on the generalized Bochner's theorem.
\end{abstract}

\section{Introduction}\label{intro}

Let us consider the following function
\begin{equation}\label{kernelLobachevskyAS}
u \times v \mapsto 
\langle u| v \rangle = \exp\Big\{ - {\textstyle\frac{\mathpzc{e}^2}{\pi}}(\lambda \textrm{coth} \lambda - 1) \Big\},
\end{equation}
for $u,v$ ranging over the Lobachevsky space $u\cdot u = 1$, $v \cdot v = 1$. 
Here $\lambda$ is the hyperbolic angle
between $u$ and $v$: $\textrm{cosh} \, \lambda = u \cdot v$. 
Although it is not evident, 
(\ref{kernelLobachevskyAS}) is equal to an invariant positive definite kernel on the Lobachevsky space. 
\cite{Staruszkiewicz1992ERRATUM} gives a decomposition of 
(\ref{kernelLobachevskyAS}) into the Fourier integral
(here $z = \mathpzc{e}^2/\pi$ and the second term below is absent for $z>1$)
\begin{multline}\label{decompositionAS}
\langle f | f\rangle = \int du dv \overline{f(u)} f(v) \langle u|v\rangle =
\frac{1}{(2\pi)^3} \int \limits_{0}^{\infty} \,
d \nu \, \nu^2 \, K(\nu; z) \, \int \limits_{\mathbb{S}^2} \, d^2 p \,
|\mathcal{F}f(p;\nu)|^2 \\ +
\frac{(1-z)^2(2e)^z}{16\pi^2} \int \limits_{\mathbb{S}^2 \times \mathbb{S}^2} 
\frac{d^2p \, d^2 k}{(p \cdot k)^z} \,\, \overline{\mathcal{F}f(p;i(1-z))} \,\,
\mathcal{F}f(k;i(1-z))
\end{multline}
for $|f\rangle = \int du f(u) |u\rangle$ with smooth $f$ of compact support on the Lobachevsky
space $u \cdot u = 1$, with the invariant measure $du$ on the Lobachevsky space. In the formula 
(\ref{decompositionAS}) the constant $e$ stands for the basis of natural logarithms and $\mathpzc{e}$ in $z = \mathpzc{e}^2/\pi$ stands for the 
elementary charge, experimental value of which is approximately equal $\tfrac{1}{\sqrt{137}}$ in units in which
$\hbar = c = 1$. Here the
Gelfand-Graev-Vilenkin inverse Fourier transform of $f$ on the Lobachevsky space is used
\[
f(u) = \frac{1}{(2\pi)^3} \int \limits_{0}^{\infty} \,
d \nu \, \nu^2 \, \int \limits_{\mathbb{S}^2} \, d^2 p \,
\mathcal{F}f(p;\nu) \, (p \cdot u)^{-i\nu-1}
\] 
together with the Gelfand-Graev-Vilenkin Fourier transform  $\mathcal{F}f$ of $f$ 
on the Lobachevsky space, equal
\[
\mathcal{F}f(p;\nu) = \int du \, f(u) \, (p \cdot u)^{i\nu-1},
\]
which is  a homogeneous of degree $i\nu-1$ function of $p$ on the positive sheet of the cone
(and thus with $\mathcal{F}f(p;i(1-z))$ homogeneous of degree $z-2$ in $p$). 
Decomposition (\ref{decompositionAS}) can be computed as in \cite{Staruszkiewicz1992ERRATUM}
without the assumption of positive definiteness of (\ref{kernelLobachevskyAS}) (the invariance of 
(\ref{kernelLobachevskyAS}) is evident).
Positive definiteness of (\ref{kernelLobachevskyAS}) is equivalent (in terms of 
\cite{Staruszkiewicz1992ERRATUM}) to the positivity of the weight function $K(\nu;z=\mathpzc{e}^2/\pi)$:
\[
K(\nu;z) = -{\textstyle\frac{4\pi}{\nu}}z^2 e^{z} 
\sum\limits_{n=-\infty}^{+\infty}{\textstyle\frac{[\nu+i(2n+1-z)]^{n-1}}{[\nu+i(2n+1+z)]^{n+2}}}
\]
in (\ref{decompositionAS}) for
each positive real $\nu$. However, positivity of the weight function 
$K(\nu;\mathpzc{e}^2/\pi)$ is not evident, compare \cite{Staruszkiewicz2020}. In fact a proof of positivity
of the kernel (\ref{kernelLobachevskyAS}), independent of the Staruszkiewicz theory \cite{Staruszkiewicz}, would give us
a proof of (relative) consistency of his theory. This is the task of the present paper.

\section{Positivity of a Hermitian form on a space of homogeneous states}\label{HermitianForm}

In this Section we define a linear space $(E^{*})_{tr}^{e}$ of generalized homogeneous states, and a Hermitian form 
$(\cdot,\cdot)_{{}_{\mathfrak{J}}}$ on this space.
We then prove its positivity. In the next Section we use positivity of $(\cdot,\cdot)_{{}_{\mathfrak{J}}}$ in the proof of positivity of the kernel
(\ref{kernelLobachevskyAS}).
Although the space $(E^{*})_{tr}^{e}$  and the form $(\cdot,\cdot)_{{}_{\mathfrak{J}}}$ have some deeper relation with
Staruszkiewicz theory, we do not enter into this relation here, and treat the space $(E^{*})_{tr}^{e}$ and the form 
$(\cdot,\cdot)_{{}_{\mathfrak{J}}}$ on $(E^{*})_{tr}^{e}$ as an intermediate auxiliary construction in the proof of positivity
of the kernel (\ref{kernelLobachevskyAS}).

We define the linear space $(E^{*})_{tr}^{e}$ of electric type transversal homogeneous of degree $\chi=-1$ states 
as the space of states spanned (over $\mathbb{C}$) by the following states 
\begin{equation}\label{SolMaxwellHom=-1}
\widetilde{f}_\mu(p) = \sum \limits_{i}^{N} \, \alpha_i {\textstyle\frac{u_{i\mu}}{u_i \cdot p}}, \,\,\,
\sum \limits_{i}^{N} \alpha_i = 0,
\end{equation}
where $u_i$ runs over a finite set of time like unit ($u_i \cdot u_i = 1$) four-vectors, 
and $p$ runs over the positive energy sheet of the cone $p\cdot p =0$ in momentum space, regarded as distribution
supported at the cone $p\cdot p =0$ in momentum $p$. 
Note that if we allow in this definition only real $\widetilde{f}$ and $\alpha_i$ and both energy sheets of the 
light cone in the momentum space, and finally discard the condition $\sum \alpha_i = 0$,
then we obtain the space of (Fourier transforms of) homogeneous of degree $-1$ solutions of d'Alembert equation
-- the \emph{electric type} solutions generated by the Dirac solution 
\[
\widetilde{f}(p) = {\textstyle\frac{u}{u \cdot p}}, \,\,\,\, u = (1,0,0,0),
\,\,\,\,\,\,\,\,\,\,\,\,
f(x) = \big(\theta(-x\cdot x) {\textstyle\frac{1}{|\boldsymbol{\x}|}},0,0,0 \big).
\]
 
Note that the condition
\[
\sum \limits_{i} \alpha_i = 0
\]
is equivalent to the transversality condition
\[
p^\mu \tilde{f}_\mu = 0,
\]
which, together with the assumption that $\supp \tilde{f}_\mu \subset \{p; p \cdot p = 0\}$) assures the inverse Fourier transform 
of $\tilde{f}_\mu$, regarded as distribution concentrated on the cone, 
to be a solution of the vacuous Maxwell equations.

For the infrared fields having the form (\ref{SolMaxwellHom=-1}) we define the invariant inner product
\begin{equation}\label{krein-prod-infra-red-1}
(\tilde{f},\tilde{f})_{{}_{\mathfrak{J}}}= \int \limits_{\mathbb{S}^2} 
\Big( \tilde{f}(p), \mathfrak{J}_{\bar{p}} \tilde{f}(p) \Big)_{\mathbb{C}^4} 
\, \ud^2 p 
\,\,\,
= 
\,\,\,
- \, \int \limits_{\mathbb{S}^2}
\overline{ 
\tilde{f}_\mu(p)
}
\tilde{f}^\mu(p)
\, \ud^2 p,
\end{equation}
\[
\mathfrak{J}_{\bar{p}} = \begin{pmatrix}
-1 & 0 & 0 & 0\\
0 & 1 & 0 & 0 \\
0 & 0 & 1 & 0 \\
0 & 0 & 0 & 1
\end{pmatrix}, 
\,\,\,\, p = (p_0,p_1, p_2, p_3) = (p_0, \boldsymbol{\p}), \,\,\, p \cdot p =0, \, p_0 >0.
\]
Recall, that by the homogeneity condition, $\widetilde{f}$ is determined by its values on the unit two sphere $\mathbb{S}^2$,
$|\boldsymbol{\p}|=1$, in the cone $\mathscr{O}_{\bar{p}} = \{p: p \cdot p = 0, p_0 > 0\}$, 
and $\ud^2 p$ can be identified with the ordinary measure on $\mathbb{S}^2$.

In case when both sheets of the light cone in momentum space are allowed, and $\widetilde{f}$ as well as 
$\alpha_i$ are real, then the sum (\ref{SolMaxwellHom=-1}) can be realized physically as the electromagnetic 
potential of the infrared radiation field produced in the scattering process of point charges
$\alpha_i$, with some four-velocities $p_{i}$ coming in  (which have, say, the corresponding $\alpha_i$ positive) and with the 
four-velocities $p_i$ coming out which have  the corresponding $\alpha_i$ with the opposite sign, 
compare \cite{Staruszkiewicz1981}. In particular for the potential 
\[
\tilde{f}_\mu(p) = \frac{\mathpzc{e}}{2\pi} \bigg( \frac{u_\mu}{u \cdot p} - \frac{v_\mu}{v \cdot p} \bigg)
\]
corresponding to the infrared field produced by a point charge $\mathpzc{e}$ scattered at the origin such that 
$u^\mu, v^\mu$ are the time like four-velocities of the point charge before and after the scattering respectively, 
the inner product (\ref{krein-prod-infra-red-1}) is equal
\[
(\tilde{f},\tilde{f})_{{}_{\mathfrak{J}}} = 2 \frac{\mathpzc{e}^2}{\pi} \bigg( \lambda \textrm{coth} \lambda - 1\bigg),
\] 
where $\lambda$ is the hyperbolic angle between $u$ and $v$, i.e. $\textrm{cosh} \, \lambda
= u\cdot v$, compare \cite{Staruszkiewicz1981}.

In the investigation of the Hermitian form (\ref{krein-prod-infra-red-1}) the operator $B$ standing
in the formula for the inner product 
$(\widetilde{\varphi}, \widetilde{\varphi}') = (\widetilde{\varphi}, B \widetilde{\varphi}')_{{}_{L^2(\mathbb{R}^3; \mathbb{C}^4)}}$ 
and in the formula for the Krein-inner product 
\[
(\widetilde{\varphi}, \mathfrak{J}'\widetilde{\varphi}') 
= (\widetilde{\varphi}, B\mathfrak{J}' \widetilde{\varphi}')_{{}_{L^2(\mathbb{R}^3; \mathbb{C}^4)}}
= (\widetilde{\varphi}, \mathfrak{J}_{\bar{p}} \widetilde{\varphi}')_{{}_{L^2(\mathbb{R}^3, \ud \mu_{\mathscr{O}_{\bar{p}}}; \mathbb{C}^4)}},
\] 
in the single particle Krein-Hilbert space $\mathcal{H}'$ of the free e.m. potential field in the Gupta-Bleuler gauge, will be useful. 
Here $\ud \mu_{\mathscr{O}_{\bar{p}}}(\boldsymbol{\p}) = \tfrac{\ud^3 \boldsymbol{\p}}{2|\boldsymbol{\p}|}$ is the standard invariant measure on the cone
$\mathscr{O}_{\bar{p}}$. Recall, that $B$ is the operator of pointwise multiplication by the matrix 
\[
\frac{1}{2 r} B(p), \,\,\, p \in \mathscr{O}_{\bar{p}};
\]
which is strictly positive and self-adjoint in $\mathbb{C}^4$, with
\[
B(p) =  \\
\left( \begin{array}{cccc} 
\frac{r^{-2} + r^2}{2} & \frac{r^{-2} - r^2}{2r}p^1 & \frac{r^{-2} - r^2}{2r}p^2 & \frac{r^{-2} - r^2}{2r}p^3 \\
\frac{r^{-2} - r^2}{2r} p^1&\frac{r^{-2} + r^2 -2}{2r^2}p^1 p^1 +1 & \frac{r^{-2} + r^2 -2}{2r^2}p^1 p^2 & \frac{r^{-2} + r^2 -2}{2r^2} p^1 p^3  \\
\frac{r^{-2} - r^2}{2r}p^2 & \frac{r^{-2} + r^2 -2}{2r^2}p^2 p^1 &\frac{r^{-2} + r^2 -2}{2r^2}p^2 p^2 +1 & \frac{r^{-2} + r^2 -2}{2r^2} p^2 p^3  \\
\frac{r^{-2} - r^2}{2r}p^3 & \frac{r^{-2} + r^2 -2}{2r^2}p^3 p^1 & \frac{r^{-2} + r^2 -2}{2r^2}p^3 p^2 &\frac{r^{-2} + r^2 -2}{2r^2}p^3 p^3 +1  \end{array}\right), 
\]
again strictly positive self-adjoint on $\mathbb{C}^4$. Here $r = p^0(\boldsymbol{\p}) = |\boldsymbol{\p}|$. 
For each $p \in \mathscr{O}_{\bar{p}}$ 
\begin{multline*}
{w_{{}_1}}^+(p) = \left( \begin{array}{c} 0 \\
                               \frac{p^2}{\sqrt{(p^1)^2 + (p^2)^2}}     \\
                             \frac{-p^1}{\sqrt{(p^1)^2 + (p^2)^2}}   \\
                               0       \end{array}\right), 
{w_{{}_1}}^- (p)= \left( \begin{array}{c} 0 \\
                               \frac{p^1 p^3}{\sqrt{(p^1)^2 + (p^2)^2}r}     \\
                             \frac{p^2 p^3}{\sqrt{(p^1)^2 + (p^2)^2}r}   \\
                              - \frac{\sqrt{(p^1)^2 + (p^2)^2}}{r}       \end{array}\right), \\ 
w_{{}_{r^{-2}}}(p) = \left( \begin{array}{c} \frac{1}{\sqrt{2}} \\
                              \frac{1}{\sqrt{2}}\frac{p^1}{r}     \\
                           \frac{1}{\sqrt{2}}\frac{p^2}{r}    \\
                             \frac{1}{\sqrt{2}}\frac{p^3}{r}       \end{array}\right),
w_{{}_{r^2}}(p) = \left( \begin{array}{c} \frac{1}{\sqrt{2}} \\
                              -\frac{1}{\sqrt{2}}\frac{p^1}{r}     \\
                           -\frac{1}{\sqrt{2}}\frac{p^2}{r}    \\
                             -\frac{1}{\sqrt{2}}\frac{p^3}{r}       \end{array}\right) 
\end{multline*} 
are the eigenvectors of the matrix $B(p)$ which are orthonormal in $\mathbb{C}^4$, where 
${w_{{}_1}}^+(p), {w_{{}_1}}^-(p)$ correspond to the eigenvalue equal $+1$, and 
$w_{{}_{r^{-2}}}(p), w_{{}_{r^2}}(p)$ correspond to the eigenvalues $r^{-2}, r^2$ respectively, compare \cite{wawrzycki2016bialynicki}.

There is a canonical decomposition of the one particle Krein-Hilbert space 
$\mathcal{H}'$ of the field $A_\mu$ associated to the operator $B$, which allows the construction 
of the subspace $\mathcal{H}'_{tr} \subset \mathcal{H}'$ of physical transversal states. 
The decomposition of $\mathcal{H}'$ associated to $B$ can in principle be extended over the space
of homogeneous functions on the cone. If in addition restrictions of these functions to the unit sphere
$\mathbb{S}^2$ belong to $L^2(\mathbb{S}^2)$, then this decomposition will allow us to make some 
statements concerning positivity of the form (\ref{krein-prod-infra-red-1}) 
as defined on homogeneous of degree $-1$
four-vector functions summable on $\mathbb{S}^2$.   

Namely, recall that the ordinary one particle state, i.e. a four-component function $\widetilde{\varphi}^\mu$ 
on the cone -- an element of the Hilbert space 
$\mathcal{H}'$, has the unique decomposition
\[
\widetilde{\varphi} = 
{w_1}^+ \tilde{f}_+ + {w_1}^- \tilde{f}_- + w_{r^{-2}} \tilde{f}_{0+} + w_{r^{2}} \tilde{f}_{0-}.
\] 
Here the four-component functions $w$, are given above, and 
are at each point $p$ of the cone $\mathscr{O}_{1,0,0,1}$ equal to the 
eigenvectors of the $4 \times 4$ matrix $B(p)$ given above. The complex valued functions $\tilde{f}_+, \tilde{f}_-$ are square integrable on the cone with respect to the invariant measure $\tfrac{\ud^3 \boldsymbol{\p}}{2|\boldsymbol{\p}|}$ on the cone, and the scalar function $\tilde{f}_{0+}$ is square integrable with respect to the measure $\frac{\ud^3 \boldsymbol{\p}}{|\boldsymbol{\p}|^3}$. Finally the complex valued function $\tilde{f}_{0-}$ is square integrable on the cone with respect to the measure $|\boldsymbol{\p}| \, \ud^3 \boldsymbol{\p}$. 
The subspace $\mathcal{H}'_{tr}$ of physical (one particle) states 
consists precisely of all those functions $\widetilde{\varphi}$ which have the decomposition
\[
\widetilde{\varphi} = {w_1}^+ \tilde{f}_+ + {w_1}^- \tilde{f}_-.
\] 
Note in particular that the elements of $\mathcal{H}'_{tr}$ are transversal in the stronger sense, i.e. not only 
$p^\mu \widetilde{\varphi}_\mu = 0$ but $p_1\widetilde{\varphi}_1 + p_1 \widetilde{\varphi}_2 + p_3 \widetilde{\varphi}_3 =0$. Let now the (four-component) function  
$\widetilde{\varphi}$ be replaced with a function $\widetilde{f}$ on the cone, homogeneous of degree $\chi =-1+i\nu$,
$\nu \in \mathbb{R}$.  In this case $\widetilde{f}$ likewise has the unique decomposition 
\[
\widetilde{f} = {w_1}^+ \tilde{f}_+ + {w_1}^- \tilde{f}_- + w_{r^{-2}} \tilde{f}_{0+} + w_{r^{2}} \tilde{f}_{0-},
\] 
where in this decomposition the functions $\tilde{f}_+, \tilde{f}_-, \tilde{f}_{0+}, \tilde{f}_{0-}$, 
are homogeneous of degree $\chi =-1+i\nu$, as the functions
${w_1}^+, {w_1}^-,w_{r^{-2}}, w_{r^{2}}$ are homogeneous of degree zero functions on the light cone. 
We assume that the functions $\widetilde{f}$ are 
regular enough in having the restrictions to the unit sphere $\mathbb{S}^2$ which 
belong to $L^2(\mathbb{S}^2)$. In this case, the decomposition of 
$\widetilde{f}$ can be used to the analysis of the positivity of (\ref{krein-prod-infra-red-1}) on the linear space of homogeneous of degree
$\chi=-1+i\nu$ states, or in particular on homogeneous of degree $-1$ states of the form (\ref{SolMaxwellHom=-1}) which can be transversal, i.e.
\[
\sum \limits_{i}^{N} \alpha_i = 0
\]
or not necessary transversal, i.e. 
\[
\sum \limits_{i}^{N} \alpha_i \neq 0.
\]
In particular
we can consistently define the physical subspace $(E^{*})_{tr}$ of homogeneous states 
as the space of all those functions on the cone
which can be represented as the linear combination
\[
{w_1}^+ \tilde{f}_+ + {w_1}^- \tilde{f}_- + w_{r^{-2}} \tilde{f}_{0+}
\]
with $\tilde{f}_+,\tilde{f}_-, \tilde{f}_{0+}$ homogeneous of degree $\chi = -1+i\nu$,
with restrictions to $\mathbb{S}^2$ belonging to $L^2(\mathbb{S}^2)$. 
Note in particular that the elements of 
$(E^{*})_{tr}$ are transversal: $p^\mu \tilde{f}_\mu = 0$.

Observe that for homogeneous transversal states $\tilde{f}_\mu$ of homogeneity degree $\chi=-1+i\nu$, or
for any $\tilde{f}_\mu \in (E^{*})_{tr}^{e}$ of the general form (\ref{SolMaxwellHom=-1})
with
\[
\sum \limits_{i}^{N} \alpha_i = 0,
\]
the bilinear form (\ref{krein-prod-infra-red-1}) is non negatively defined. Indeed 
any such element can be decomposed into the three components
\[
\tilde{f}_\mu = {{w_1}^+}_\mu \tilde{f}_+ + {{w_1}^-}_\mu \tilde{f}_- + {w_{r^{-2}}}_\mu \tilde{f}_{0+},
\]
(the fourth component of the general decomposition is lacking because of the transversality).
On the other hand, the components ${{w_1}^+}_\mu \tilde{f}_+$, ${{w_1}^-}_\mu \tilde{f}_-$, 
${w_{r^{-2}}}_\mu \tilde{f}_{0+}$ are orthogonal with respect to (\ref{krein-prod-infra-red-1}),
the bilinear form (\ref{krein-prod-infra-red-1}) is positive (for the first two components 
${{w_1}^+}_\mu \tilde{f}_+$, ${{w_1}^-}_\mu \tilde{f}_-$)
or zero (for the last ${w_{r^{-2}}}_\mu \tilde{f}_{0+}$). Thus non negativity on transversal
homogeneous of degree $\chi=-1+i\nu$ as well as positivity
on the states (\ref{SolMaxwellHom=-1})
follows whenever 
\[
\sum \limits_{i}^{N} \alpha_i = 0.
\]

Thus we may summarize the results in the following
\begin{lem*}
The invariant Hermitian bilinear form (\ref{krein-prod-infra-red-1})  
\[
(\tilde{f},\tilde{f})_{{}_{\mathfrak{J}}}= \int \limits_{\mathbb{S}^2} 
\Big( \tilde{f}(p), \mathfrak{J}_{\bar{p}} \tilde{f}(p) \Big)_{\mathbb{C}^4} 
\, \ud^2 p 
\,\,\,
= 
\,\,\,
- \, \int \limits_{\mathbb{S}^2} 
\overline{
\tilde{f}_\mu(p)
}
\tilde{f}^\mu(p)
\, \ud^2 p
\]
is non-negatively definite on the linear space  $(E^{*})_{tr}$ of transversal homogeneous of degree
$\chi = -1+i\nu$ states as well as on the space $(E^{*})_{tr}^{e}$ of transversal electric-type 
states
\[
\tilde{f}_\mu(p) = \sum \limits_{i}^{N} \, \alpha_i \frac{u_{i\mu}}{u_i \cdot p}, \,\,\,\,\,\,\,
\sum \limits_{i}^{N} \alpha_i = 0, \,\, u_i \cdot u_i =1, \, i=1, \ldots, N.
\]
\end{lem*}

\section{Positivity of the invariant kernel $\langle \cdot|\cdot\rangle$ on the Lobachevsky space}\label{kernel}

Now we construct a continuous and invariant kernel $\langle \cdot|\cdot\rangle$ 
on the Lobachevsky space $\mathscr{L}_3$,
which is of considerable importance for Staruszkiewicz theory. We give a proof of its positivity. 
In order to achieve this result, we use the linear subspace $L[\mathfrak{F}_{\chi=1}]$ 
of states homogeneous of degree $-1$.

For this reason, consider now the specific homogeneous of degree $-1$ state (Fourier transform
of the Dirac homogeneous of degree $-1$ solution restricted to the positive energy sheet of the cone)
of the form
\begin{equation}\label{tildefu}
\widetilde{f}^{|u\rangle}_{\mu}(p) = \frac{u_\mu}{u \cdot p}
\end{equation}
with a fixed unit time like vector $u$ in the Lobachevsky space. 
Then construct the linear span $L[\mathfrak{F}_{\chi=1}]$ of all such (\ref{tildefu}) with $u$
ranging over the Lobachevsky space $\mathscr{L}_3$ of unit time like vectors $u$, $u\cdot u =1$.
In other words, we consider the space $L[\mathfrak{F}_{\chi=1}]$
spanned by all Lorentz   
transforms
\[
\begin{array}{cc}
\Lambda(g)^{-1} \tilde{f}^{|u\rangle}(\Lambda(g)p) =
\tilde{f}^{|u'\rangle}(p) = \frac{u'}{u' \cdot p}, \,\,\,
u' = \Lambda(g)^{-1} u,
\\
g \in SL(2, \mathbb{C}),
\end{array}
\]
of one single  state of the form $\tilde{f}^{|u\rangle}$. Here,  
we have used the natural antihomomorphism $g \rightarrow \Lambda(g)$ 
from the $SL(2, \mathbb{C})$ into the group of Lorentz transformations.

Note that this Lorentz transformation is induced by the linear dual of the 
(conjugated) {\L}opusza\'nski transformation acting in the Fock space of the quantum field $A_\mu$.  
Namely, we put the natural formula 
for the invariant pairing
\begin{multline}\label{pairing-formula}
(\tilde{f}, \widetilde{\varphi})_{{}_{\textrm{pairing}}} = (\tilde{f}, \mathfrak{J}'\widetilde{\varphi})
= - \int \limits_{\mathscr{O}_{\pm 1,0,0,1}} \tilde{f}^\mu (p) \widetilde{\varphi}_\mu (p) \, \ud \mu_{{}_{\mathscr{O}_{\pm 1,0,0,1}}}(p) \\
= \int \limits_{\mathscr{O}_{\pm 1,0,0,1}} \big( \tilde{f} (p), \mathfrak{J}_{{}_{\bar{p}}} 
\widetilde{\varphi} (p) \big)_{{}_{\mathbb{C}^4}} \, \ud \mu_{{}_{\mathscr{O}_{\pm 1,0,0,1}}}(p).
\end{multline}

The space $L[\mathfrak{F}_{\chi=1}]$ contains the transversal electric type homogeneous states (respectively homogeneous of degree $-1$
solutions of d'Alembert equation) of the form
(\ref{SolMaxwellHom=-1}) with
\[
\sum \limits_{i}^{N} \alpha_i = 0
\]
along with the longitudinal solutions of the form (\ref{SolMaxwellHom=-1}) with
\[
\sum \limits_{i} \alpha_j \neq 0.
\]

The space of invariant kernels $\langle \cdot | \cdot \rangle$ on $\mathscr{L}_3$ is rather reach. 
However, in this particular case of positive definite kernels on the Lobachevsky space 
$\mathscr{L}_3 \cong SL(2, \mathbb{C})/SU(2, \mathbb{C})$ acted on by $SL(2, \mathbb{C})$
the invariant kernels are fully classified, compare e.g. \cite{Gangolli}.
The manifold $\mathscr{L}_3$ can also be realized by $2\times 2$ Hermitian complex matrices $\widehat{u} = \sigma_\mu u^\mu$, where 
$\sigma_0 = \boldsymbol{1}_{{}_{2}}$ and $\sigma_i$, $i=1$,$2$,$3$, are the Pauli matrices and $u\cdot u =1$. Next we consider the smooth 
left action $SL(2, \mathbb{C}) \times \mathscr{L}_3 \ni g \times \widehat{u} \rightarrow g \cdot \widehat{u} \in \mathscr{L}_3$
of  $SL(2, \mathbb{C})$ on $\mathscr{L}_3$ defined by the formula
\[
g \cdot \widehat{u} = g \widehat{u} g^* = \widehat{\Lambda(g^{-1})u}, \,\,\,\,\,\,\,\,\,
g \in SL(2, \mathbb{C}), \widehat{u} \in \mathscr{L}_3.
\]
Then  $\mathscr{L}_3$ is equal to the orbit $\mathscr{O}_{{}_{\widehat{u}}} = \{g \cdot \widehat{u}, g \in  SL(2, \mathbb{C}) \}$ of
the point $\widehat{u}$ with $u = (1,0,0,0)$. The isotropy group at the point $\widehat{u}$ with $u = (1,0,0,0)$, is equal
to the maximal compact subgroup  $K = SU(2, \mathbb{C})$ of $SL(2, \mathbb{C})$. Therefore, $\mathscr{L}_3 = \mathscr{O}_{{}_{\widehat{u}}}$,
$u = (1,0,0,0)$, is diffeomorphic to the space $\mathscr{L}_3 \cong SL(2, \mathbb{C})/SU(2, \mathbb{C})$ of left cosets
$qK$, $q \in SL(2, \mathbb{C})$, with the above left action, which coincides, under this diffeomorphism, with the ordinary left action
$g\times qK \rightarrow gqK$ on the left cosets $qK$. To the coset $\mathfrak{e}K$ of the identity element $\mathfrak{e} \in SL(2, \mathbb{C})$
there corresponds the point $\widehat{u}$, $u = (1,0,0,0)$, invariant under $SU(2, \mathbb{C})$.
Choosing various invariant positive definite kernels $\langle \cdot | \cdot \rangle$ on $\mathscr{L}_3$ we achieve in this way various cyclic spherical unitary 
representations $\mathbb{U}$ of the $SL(2, \mathbb{C})$ group
on the completion of $L[\mathfrak{F}_{\chi=1}]$ with respect to the inner product defined by the invariant kernel.  Let $K= SU(2, \mathbb{C})$ be the maximal compact subgroup of 
$SL(2, \mathbb{C})$. Let a unitary representation $U$ of $SL(2, \mathbb{C})$ be called $K$-spherical (or merely spherical) if the decomposition
of the restriction of $U$ to $K$ contains the trivial representation $k \rightarrow 1$ of $K$. Equivalently $U$
is spherical whenever there is a unit vector $v \in H_{U}$ such that $U_k v = v$ for all $k \in K$.
Then in particular it follows by the classification results (or Gelfand's theory of spherical functions
and his generalization of Bochner's theorem for semi-simple Lie groups, in particular for $SL(2, \mathbb{C})$ group), \cite{Gangolli}, that each unitary cyclic and spherical representation $\mathbb{U}$ of $SL(2, \mathbb{C})$ can be reached by the respective choice of the invariant kernel on the Lobachevksy space, or to each such representation there exists the corresponding invariant kernel. 

It follows that the most general representation $\mathbb{U}$ which can be achieved in this way
has the general form \cite{Gangolli}:
\begin{equation}\label{BochnerKernelDecomposition}
\mathbb{U} = \int \limits_{\mathbb{R}} \mathfrak{S}(m = 0, \rho) \, \ud \rho \oplus \int \limits_{[0,1] \subset \mathbb{R}} \mathfrak{D}(\nu) \, \ud \nu
\end{equation}
where $\mathfrak{S}(m , \rho)$ is the irreducible representation of the principal series denoted by the pair
$(l_0 = \frac{m}{2}, l_1 = \frac{i\rho}{2})$, with $m \in \mathbb{Z}$
and $\rho \in \mathbb{R}$ in the notation of the book \cite{Geland-Minlos-Shapiro}, and correspond to the characters
$\chi = (n_1, n_2) = \big(\frac{m}{2} + \frac{i\rho}{2}, - \frac{m}{2} + \frac{i\rho}{2}\big)$ in the notation
of the book \cite{GelfandV}. Here $\mathfrak{D}(\nu)$
are the irreducible unitary representations of the supplementary series denoted by the pair
$(l_0 = 0, l_1 = \nu)$ in the notation of the book \cite{Geland-Minlos-Shapiro}, and correspond to the character
$\chi = (n_1, n_2) = \big(\nu, \nu)$ in the notation
of the book \cite{GelfandV} with\footnote{In the notation 
of \cite{nai1}-\cite{nai3} the parameter $\nu$ numbering the supplementary sries $\mathfrak{D}(\nu)$
is twice as ours $\nu$ and ranges over the interval $(0,2)$.} the real parameter $\nu \in (0,1)$. Finally 
$\ud \rho$ and $\ud \nu$ are arbitrary $\sigma$-measures 
on the reals $\mathbb{R}$ and on the interval $[0,1] \subset \mathbb{R}$ respectively. 

However the classification of positive definite invariant kernels on the Lobachevsky space, as presented
e.g. in \cite{Gangolli}, requires a considerable work in each particular case, needed to 
give a more concrete form to the possible kernels, compare e.g. the example of positive definite kernels on the Lobachevsky plane 
\[
\mathscr{L}_2 =  SL(2, \mathbb{R})/SO(2)
\]
invariant under $SL(2, \mathbb{R})$. Unfortunately the case 
\[
\mathscr{L}_3 =  SL(2, \mathbb{C})/SU(2)
\]
 has not been worked out 
in \cite{Gangolli} in explicit form. Therefore we prefer to construct the required kernel, which is of particular importance, with the help of the Hermitian form (\ref{krein-prod-infra-red-1}).

Recall that for
two points $u, v$ of the Lobachevsky space we have
\[
(\tilde{f}^{|u\rangle},\tilde{f}^{|v\rangle})_{{}_{\mathfrak{J}}} = -4\pi \lambda \textrm{coth} \lambda,
\]
where $\lambda$ is the hyperbolic angle between $u$ and $v$: $\textrm{cosh} \, \lambda = u\cdot v$, 
compare \cite{Staruszkiewicz1981}. The Hermitian bilinear invariant form (\ref{krein-prod-infra-red-1}) is not positive definite on the linear space $L[\mathfrak{F}_{\chi=1}]$ of states spanned by the states 
$\tilde{f}^{|u'\rangle}$ of the form (\ref{tildefu}) with $u'$
ranging over the Lobachevsky space $\mathscr{L}_3$. Nonetheless it defines (after addition of the constant term
$4\pi$ and changing the sign) the ``polarization'' of a L\'evy-Schoenberg kernel on the Lobachevsky space
$\mathscr{L}_3 =  SL(2, \mathbb{C})/K = SL(2, \mathbb{C})/SU(2, \mathbb{C})$ (we are using the terminology of 
\cite{Gangolli}).
Namely the kernel 
\[
u \times v \mapsto -((\tilde{f}^{|u\rangle},\tilde{f}^{|v\rangle})_{{}_{\mathfrak{J}}} + 4\pi)
\]
on $\mathscr{L}_3 = SL(2, \mathbb{C})/SU(2, \mathbb{C})$ preserves the conditions 
(2.16)-(2.19) of \cite{Gangolli}. 
In particular (2.19) of \cite{Gangolli} means in our case that for each positive real number $t$ 
\begin{equation}\label{kernel-t}
u \times v \mapsto \langle u|v\rangle_t = e^{t((\tilde{f}^{|u\rangle},\tilde{f}^{|v\rangle})_{{}_{\mathfrak{J}}} + 4\pi)}
=
e^{-t4\pi(\lambda \textrm{coth}\lambda -1)}
\end{equation} 
is an invariant positive definite kernel 
on the Lobachevsky space  and thus defines positive definite 
and invariant inner product on the linear space $S$ spanned by $\tilde{f}^{|u\rangle}$
and all its Lorentz transforms $\tilde{f}^{|u'\rangle}$ defined by (\ref{tildefu}) with $u'$ ranging over 
the Lobachevsky space. Here $\lambda$ is the hyperbolic angle between $u$ and $v$.

Indeed that the conditions (2.16)-(2.18) of \cite{Gangolli} are preserved is immediate. We need only show that 
(2.19) of \cite{Gangolli} is preserved, i.e. that the kernel (\ref{kernel-t})
is positive definite. But in order to see this note that
\[
\Big( \sum \limits_{i} \alpha_i \tilde{f}^{|u_i \rangle}, \sum \limits_{j} \alpha_j \tilde{f}^{|u_j\rangle} 
\Big)_{{}_{\mathfrak{J}}} \geq 0
\]
whenever 
\[
\sum \limits_{i} \alpha_i =0
\]
for $\tilde{f}^{|u\rangle}$ defined by (\ref{tildefu}), as we have already shown that the bilinear form 
$(\cdot,\cdot)_{{}_{\mathfrak{J}}}$ is positive definite on the linear space of electric type transversal states
(\ref{SolMaxwellHom=-1}), compare the Lemma of Section \ref{HermitianForm}.
This means that the function
\[
u \times v \mapsto -((\tilde{f}^{|u\rangle},\tilde{f}^{|v\rangle})_{{}_{\mathfrak{J}}} + 4\pi)
\]
is a conditionally negative definite kernel on the Lobachevsky space in the sense of Schoenberg
\cite{Schoenberg}, compare also \cite{PaulsenRaghupathi}, \S 9.1. Thus by the classical result 
of Schoenberg \cite{Schoenberg}  (compare e. g. also
\cite{PaulsenRaghupathi}, \S 9.1, Theorem 9.7) 
\[
u \times v \mapsto \langle u|v\rangle_t = e^{t((\tilde{f}^{|u\rangle},\tilde{f}^{|v\rangle})_{{}_{\mathfrak{J}}} + 4\pi)}
=
e^{-t4\pi(\lambda \textrm{coth}\lambda -1)}
\]
is a positive definite kernel on the Lobachevsky space for all positive $t$. Its invariance follows from the invariance of the bilinear form $(\cdot,\cdot)_{{}_{\mathfrak{J}}}$ and the transformation rule for 
$\tilde{f}^{|u\rangle}$ defined by (\ref{tildefu}). 

This positivity result is of particular importance in the theory of Staruszkiewicz, so we state it as 
a separate
\begin{twr*}
For each positive real number $t$ the function 
\[
u \times v \mapsto \langle u|v\rangle_t = e^{t((\tilde{f}^{|u\rangle},\tilde{f}^{|v\rangle})_{{}_{\mathfrak{J}}} + 4\pi)}
=
e^{-t4\pi(\lambda \textrm{coth}\lambda -1)}
\]
defines a positive definite invariant kernel on the Lobachevsky space $\mathscr{L}_3$
of unit time like vectors $u$. Here $\lambda$ is the hyperbolic angle between $u$ and $v$
in $\mathscr{L}_3$. 
\end{twr*}

We can choose $u \times v \mapsto \langle u|v\rangle_t$ as the invariant kernel defining the inner product
$\langle \cdot | \cdot \rangle_{{}_{\mathfrak{J}, t}}$ on the 
linear space $L[\mathfrak{F}_{\chi=1}]$ of states spanned by  $\tilde{f}^{|u'\rangle}$
defined by (\ref{tildefu}), with $u'$ ranging over the Lobachevsky space, by the formula
\begin{equation}\label{krein-prod-infra-red-long}
\Big\langle \sum \limits_{i=1}^{m} \alpha_i \tilde{f}^{|u_i\rangle} \Big| 
\sum \limits_{i=1}^{m} \beta_j \tilde{f}^{|v_j\rangle} \Big\rangle_{{}_{\mathfrak{J}, t}}
= \sum \limits_{i,j=1}^{m} \overline{\alpha_i} \beta_j \langle u_i | v_j \rangle_t,
\end{equation} 
and define the Hilbert space completion $\mathcal{H}_{t} \nsubseteq E^{*}$ of it.
Then we recover the unitary representation $\mathbb{U}^{t}$ of the $SL(2, \mathbb{C})$ group which the 
action of the dual of the (conjugate) of the {\L}opusza\'nski representation induces on the  linear
space  $L[\mathfrak{F}_{\chi=1}]$  of states and its Hilbert space completion $\mathcal{H}_{t}$. 
Indeed by comparing this construction with the result  of \cite{Staruszkiewicz1992ERRATUM} 
and \cite{Staruszkiewicz2009} we obtain the following formula
\begin{equation}\label{decASlong}
\mathbb{U}^{t} = 
\left\{ \begin{array}{lll}
\mathfrak{D}(\nu_0) \bigoplus \int \limits_{\rho>0} \mathfrak{S}(m=0, \rho) \ud \rho,
& \nu_0 = 1 - 4 \pi t, & \textrm{if} \, 0 < 4 \pi t <1  \\
\int \limits_{\rho>0} \mathfrak{S}(m=0, \rho) \, \ud \rho, & &
\textrm{if} \, 1 < 4 \pi t, 
\end{array} \right.
\end{equation}
where $\ud \rho$ is the ordinary Lebesgue measure on $\mathbb{R}_+$, compare
(\ref{decompositionAS}).

\section{Positivity of the weight function $K(\nu;z)$}\label{weight}

Now let us go back to the consistency of the axioms of Staruszkiewicz theory \cite{Staruszkiewicz}.
This consistency is equivalent to the existence (proved by explicit construction) of the unitary representation $U$
of the $SL(2, \mathbb{C})$ which makes the phase field $S(x)$ a scalar field on the de Sitter hyperplane. 
Construction of this $U$ is based (compare \cite{Staruszkiewicz1995}, \cite{Staruszkiewicz1992ERRATUM}, \cite{wawrzycki-alfa}) 
on the fact that the mapping (\ref{kernelLobachevskyAS}) is equal to an invariant 
positive definite kernel on the Lobachevsky space. Positivity of (\ref{kernelLobachevskyAS}) would of course 
immediately follow from the consistency of Staruszkiewicz theory \cite{Staruszkiewicz},
but of course to show the consistency we should prove that (\ref{kernelLobachevskyAS})
is an invariant positive definite kernel independently of the consistency assumption.
But, as we have shown in Section \ref{kernel}, the function (\ref{kernelLobachevskyAS}) 
defines indeed an invariant positive definite kernel on the Lobachevsky space, 
using Schoenberg's theorem on conditionally negative definite functions. Thus, the consistency
of the axioms of \cite{Staruszkiewicz}, is thereby proved.

But, also positivity of $K(\nu;\mathpzc{e}^2/\pi)$ follows from the positive definiteness of the 
invariant kernel (\ref{kernelLobachevskyAS}) -- \emph{i.e.} from theorem of Section \ref{kernel}.  
Indeed, it follows on application of the generalization of Bochner's theorem\footnote{Due to Krein, Naimark and Gelfand \cite{Neumark_dec}, \S 31.10.},
extended to the relation between positive measures on the set of irreducible $K$-spherical unitary representations 
of semi-simple Lie groups $G$ and the corresponding positive definite kernels on $G/K \times G/K$
(or positive definite functions
on $G$, corresponding to positive definite kernels on $G/K \times G/K$), compare e.g. \cite{Gangolli}, 
with $G= SL(2, \mathbb{C})$, $K = SU(2, \mathbb{C})$ and with the Lobachevsky space $G/K = \mathscr{L}_3$
as the homogeneous Riemannian manifold. Recall, please, that a continuous function $\varphi$ on $G$ is called positive definite iff
\[
\sum_{i,j} \varphi(g_{i}^{-1}g_{j}) \alpha_i \overline{\alpha_j} \geq 0
\]
for any finite set of complex numbers $\alpha_i$, and any finite set of $g_i \in G$; and that $\varphi$ is $K$-spherical
iff $\varphi(k_1gk_2) = \varphi(g)$, for all $g \in G$, $k_1,k_2 \in K\subset G$. $\varphi$ is called normalized iff
$\varphi(\mathfrak{e})=1$. Recall, that a unitary representation $U$ of $G$ is called $K$-spherical if there exists a unit vector $v\in H(U)$,
such that $U(k)v= v$ for each $k\in K$. Recall further that to any $K$-spherical
unitary representation $U$ of $G$ there corresponds the $K$-spherical continuous normalized and positive definite
function $\varphi(g) = (U(g)v,v)$, where $(\cdot, \cdot)$ is the inner product in the Hilbert space $H(U)$ of the representation $U$. 
If the $K$-spherical representation $U$ is cyclic, with the cyclic vector $v$ which is invariant under $U(k)$, $k\in K$,
then the correspondence between the unitary equivalence class of $U$ and the spherical function $\varphi(g) = (U(g)v,v)$
is bi-unique. This is in particular the case for irreducible $K$-spherical $U$. 
If the $K$-spherical representation $U$ is irreducible, then
the corresponding  $K$-spherical normalized and positive definite
function $\varphi$ is called \emph{elementary}. Recall that each positive definite kernel $\kappa$ on $G/K \times G/K$
can be lifted to the bi-uniquelly corresponding positive definite function $\varphi$ on $G$ through the formula
$\varphi(g)= \kappa(gK, \mathfrak{e}K)$ (compare e.g. \cite{Gangolli}). Let $K$ be maximal compact subgroup of a semisimple $G$. 
Finally, let $\nu \in \mathscr{M}$ be all equivalence classes of 
all irreducible $K$-spherical representations $U_{{}_{\nu}}$ of a semisimple Lie group $G$
and let $\varphi_{{}_{\nu}}$ be the elementary normalized positive definite $K$-spherical
functions corresponding to the irreducible representants $U_{{}_{\nu}}$, one for each unitary equivalence class. 
Then we have the following generalization of Bochner's theorem
(compare \S 31.10, eq. (4), p. 426 of \cite{Neumark_dec}, or Theorem 3.23 in \cite{Gangolli}):
\begin{twr*}[{\bf Generalized Bochner's theorem}]
Let $\varphi$ be a continuous positive definite $K$-spherical function on $G$. Then
there exists a unique nonnegative measure $\mu$ on $\mathscr{M}$ such that
\[
\varphi(g) = \int\limits_{\mathscr{M}} \varphi_{{}_{\nu}}(g) \, d\mu(\nu),
\,\,\,\,\,\,\,\,\,\,\,\,\,\,\,\, g \in G.
\]
\end{twr*}

The Fourier transform of the kernel (\ref{kernelLobachevskyAS}) found in \cite{Staruszkiewicz1992ERRATUM} 
(or decomposition  (\ref{BochnerKernelDecomposition})
of the representation corresponding to the kernel (\ref{kernelLobachevskyAS}) or equivalently to the kernel (\ref{kernel-t}) with $4\pi t$ put equal 
$\mathpzc{e}^2/\pi$) 
applied to the spherical positive definite function $\varphi$ corresponding to the kernel (\ref{kernelLobachevskyAS}) gives decomposition 
of $\varphi$ into the elementary spherical functions $\varphi_{{}_{\nu}}$ with the same measure $\mu$ on the set of equivalence classes of spherical 
unitary irreducible representations of $G=SL(2, \mathbb{C})$ as in (\ref{decompositionAS}). 
On the set of spherical representations
of the principal series
it is given by the Lebesgue mesure $d\nu$ on $\mathbb{R}_+$ with the weight function equal 
$(2\pi^2)^{-1}\nu^2 K(\nu;\mathpzc{e}^2/\pi)$: $d\mu(\nu) = (2\pi^2)^{-1} \nu^2 K(\nu;\mathpzc{e}^2/\pi) d\nu$. 
Therefore, the generalized Bochner's theorem  (eq. (4), p. 426 of \cite{Neumark_dec} or Theorem 3.23 of \cite{Gangolli}) implies
positivity of the measure $d\mu(\nu) = (2\pi^2)^{-1} \nu^2 K(\nu;\mathpzc{e}^2/\pi) d\nu$. 
Thus, positivity of the weight function
$K(\nu;\mathpzc{e}^2/\pi)$ in (\ref{decompositionAS}) follows for almost all $\nu$. By the analyticity of
$K(\nu;\mathpzc{e}^2/\pi)$ in both arguments (compare \cite{Staruszkiewicz1992ERRATUM}) positivity of  
$K(\nu;\mathpzc{e}^2/\pi)$ in ordinary sense follows. Let us explain it in more detail. 
Recall, please, that the unitary irreducible representations of the principal series $\mathfrak{S}(m=0, \nu)$, $\nu \in \mathbb{R}$ and 
of the supplementary series $\mathfrak{D}(\nu_0)$, $\nu_0 \in [0,1]$, exhaust all irreducible equivalence classes of all spherical unitary irreducible
representations of $G= SL(2, \mathbb{C})$. Next, recall that the Fourier decomposition
of the kernel $\kappa(u,v) = \langle u | v \rangle$, given by (\ref{kernelLobachevskyAS}), and found in \cite{Staruszkiewicz1992ERRATUM},
is equal
\begin{multline}\label{kerneldecompositionAS}
\langle u | v \rangle = {\textstyle\frac{1}{(2\pi)^3}} \int \limits_{0}^{\infty} \,
d \nu \, \nu^2 \, K(\nu; z) \, \int \limits_{\mathbb{S}^2} \, d^2 p \,
\overline{(p\cdot u)^{i\nu-1}}(p\cdot v)^{i\nu-1} \\ +
{\textstyle\frac{(1-z)^2(2e)^z}{16\pi^2}} \int \limits_{\mathbb{S}^2 \times \mathbb{S}^2} 
\frac{d^2p \, d^2 k}{(p \cdot k)^z} \,\, \overline{(p\cdot u)^{z-2}} \,\,
(k\cdot v)^{z-2} 
\\
= \kappa(u,v) = \kappa(gK, \mathfrak{e}K) =  \varphi(g),
\\
u = gv = \Lambda(g^{-1})v, \,\,\, v= (1,0,0,0), \,\,\, g \in SL(2, \mathbb{C}), 
\end{multline}
here with $e$ in front of the second integral equal to the basis of natural logarithms and with $\mathfrak{e}$ equal to the unit
in $SL(2, \mathbb{C})$.
Recall that the $d^2 p$ integral over $\mathbb{S}^2$ in the first summand is equal to the inner product of two functions
\[
p \rightarrow (p\cdot u)^{i\nu-1}, \,\,\,\, p \rightarrow (p\cdot v)^{i\nu-1}
\]
homogeneous of degree $i\nu -1$ on the positive cone $p\cdot p =0$, $p_0>0$, in the Hilbert 
space of the unitary spherical representation of the principal series $\mathfrak{S}(m=0, \nu)$, realized as the closure
with this inner product of all homogeneous of degree $i\nu -1$ continuous functions on the cone.   
The double integral $d^2 p \times d^2 k$ over $\mathbb{S}^2 \times \mathbb{S}^2$ of the second summand in (\ref{kerneldecompositionAS})
is equal  to the inner product of two functions
\[
p \rightarrow (p\cdot u)^{z-2}, \,\,\,\, p \rightarrow (p\cdot v)^{z-2}
\]
homogeneous of degree $z-2$ on the positive cone $p\cdot p =0$, $p_0>0$, in the Hilbert 
space of the unitary spherical representation of the supplementary series $\mathfrak{D}(\nu_0 = 1-z)$, realized as the closure
with this inner product of all homogeneous of degree $2-z$ continuous functions on the cone. It is easily seen that
the function  
\[
p \rightarrow {\textstyle\frac{1}{\sqrt{4\pi}}} (p\cdot v)^{i\nu-1},  \,\,\,\, v= (1,0,0,0),
\]
represents the unit state invariant under the action of all elements of $K=SU(2, \mathbb{C})$ in the Hilbert space of 
the $K$-spherical irreducible representation $\mathfrak{S}(m=0, \nu)$ of the principal series, in the realization stated above. Similarly, 
\[
p \rightarrow \left({\textstyle\frac{(1-z)2^z}{16 \sqrt{2} \pi^{5/2}}}\right)^{1/2} (p\cdot v)^{z-2},  \,\,\,\, v= (1,0,0,0),
\]
is the function representing the unit state invariant under the action of all elements of  $K=SU(2, \mathbb{C})$ 
in the Hilbert space of the $K$-spherical representation
$\mathfrak{D}(\nu_0 = 1-z)$ of the supplementary series, in the realization stated above. Therefore, the elementary positive definite 
and normalized spherical functions $\varphi_{{}_{\nu}}$ and $\varphi_{{}_{\nu_0}}$ corresponding, respectively, to the spherical representations
of the classes $\mathfrak{S}(m=0, \nu)$ and $\mathfrak{D}(\nu_0 = 1-z)$ are, respectively, equal
\[
\varphi_{{}_{\nu}}(g) = 
{\textstyle\frac{1}{4\pi}}
\int \limits_{\mathbb{S}^2} \, d^2 p \,
\overline{(p\cdot u)^{i\nu-1}}(p\cdot v)^{i\nu-1}, \,\,\, v=(1,0,0,0), \, u = gv = \Lambda(g^{-1})v
\]
\begin{multline*}
\varphi_{{}_{\nu_0}}(g) =
{\textstyle\frac{(1-z)2^z}{16 \sqrt{2} \pi^{5/2}}} \int \limits_{\mathbb{S}^2 \times \mathbb{S}^2} 
\frac{d^2p \, d^2 k}{(p \cdot k)^z} \,\, \overline{(p\cdot u)^{2-z}}(k\cdot u)^{z-2},
\\
\,\,\, v=(1,0,0,0), \, u = gv = \Lambda(g^{-1})v.
\end{multline*}
Inserting these formulas into (\ref{kerneldecompositionAS})
we obtain
\begin{multline}\label{decompositionOfvarphi}
\varphi(g)
= \int\limits_{\mathbb{R_+}} \varphi_{{}_{\nu}}(g) \ud \mu(\nu) + \int\limits_{[0,1]} \varphi_{{}_{\nu_0}}(g) d\mu(\nu_0)
\\
= {\textstyle\frac{1}{2\pi^2}} \int \limits_{0}^{\infty} \, \varphi_{{}_{\nu}}(g) \,
\, \nu^2 \, K(\nu; z) d\nu  +
{\textstyle\frac{(1-z)e^z}{\sqrt{2\pi}}} \, \varphi_{{}_{\nu_0}}(g),
\end{multline}
with the restriction of the measure $d\mu(\nu)$ to the set $\mathbb{R}_+$ of the equivalence classes $\nu\in \mathbb{R}_+$ 
of the spherical representations of the principal series 
$\mathfrak{S}(m=0, \nu)$ equal 
\[
d\mu(\nu) = {\textstyle\frac{1}{2\pi^2}} \nu^2 \, K(\nu; z) d\nu 
\]
and with the restriction of the measure $d\mu(\nu_0)$ to the set $[0,1]$ of the equivalence classes $\nu_0\in[0,1]$ 
of the spherical representations of the supplementary series 
$\mathfrak{D}(\nu_0)$ equal to the discrete measure concentrated at the single point $\nu_0 = 1-z$, and giving
the measure 
\[
{\textstyle\frac{(1-z)e^z}{\sqrt{2\pi}}}
\] 
to the single point set $\{ \nu_0 = 1-z \}$. Recall, that the second discrete term 
in (\ref{decompositionOfvarphi}) is present if and only if $0< z <1$. In (\ref{decompositionOfvarphi}) and in the last formula, 
$e$ is equal to the basis of natural logarithms.
Thus, from the generalized Bochner theorem positivity of the weight function 
$\nu \rightarrow  K(\nu; z)$ follows for all $z, \nu >0$.

Thus, using theorem of Section \ref{kernel} and the generalized Bochner's theorem, we have proved
\begin{twr*}
The weight function $\nu \rightarrow K(\nu; z=\mathpzc{e}^2/\pi)$ in the Fourier decomposition (\ref{decompositionAS})
is non-negative.  
\end{twr*}

Let us emphasize that our proof is independent of the axioms of \cite{Staruszkiewicz}, thereby proving (relative) consistency of the axioms of \cite{Staruszkiewicz},
\emph{i.e.} we have proved that the consistency of the axioms of Staruszkiewicz theory \cite{Staruszkiewicz} is equivalent to the consistency of the harmonic analysis
for the semisimple Lie group $SL(2, \mathbb{C})$.   

One can find out that there exists a small positive $\epsilon$ such that
$\nu \rightarrow K(\nu; z=\mathpzc{e}^2/\pi)$ is a non-negative function for $\mathpzc{e}^2/\pi$ in the interval $0< \mathpzc{e}^2/\pi < \epsilon$, 
on using explicit inspection, compare \cite{Staruszkiewicz2020}. Thus, in this asymptotic case  $\mathpzc{e}^2/\pi \ll 1$,
positivity of the weight $K$ follows by explicit inspection, and then positivity
of the kernel (\ref{kernelLobachevskyAS}) for $\mathpzc{e}^2/\pi \ll 1$, by the generalized Bochner's theorem. Note here that 
the experimental value of $\mathpzc{e}^2/\pi$ is $\approx 0.0023$. Numerical calculations \cite{Staruszkiewicz2020} 
confirm positivity of the weight $K(\nu; \mathpzc{e}^2/\pi)$.

\section*{Acknowledgments}

The author is indebted for helpful discussions to prof. A. Staruszkiewicz. The author wishes to acknowledge 
the support of the Bogoliubov Laboratory of Theoretical Physics,
Joint Institute for Nuclear Research, 141980 Dubna.



\end{document}